# Polygonal walls and the astronomical alignments of the Acropolis of Alatri, Italy: a preliminary investigation.

Giulio Magli
Dipartimento di Matematica del Politecnico di Milano
P.le Leonardo da Vinci 32, 20133 Milano, Italy.

## 1 Introduction: polygonal walls in Italy

The so-called *cyclopean* or *polygonal* walls are huge walls of megalithic blocks cut in polygonal shapes and fitted together without the use of mortar. Many walls of this kind were constructed during the Bronze Age in the Mediterranean area (probably the most famous are the walls of Mycenae).[1] Slightly less known, but equally impressive and magnificent, are the walls visible in many Italian towns, spread into a wide area which spans from Umbria to Campania. Besides Alatri, which will be the subject of the present paper, those of Segni, Ferentino, Norba, Arpino and Circei are worth mentioning.

Polygonal walls are traditionally classified into three types or *manners*. The first type is constructed with blocks of medium dimensions with coarse and inaccurate joints, the second type shows a good stonework with few wedge fillings between main blocks, and finally the third manner is simply perfect, with joints so accurately prepared that it is impossible to insert even a sheet of paper or the blade of a pocket knife between two adjacent stones. The small town of Alatri, in the Frosinone provincial district at one hour driving from the center of Rome, is the place where this "third manner" of polygonal walls is still visible nearly intact, as it was constructed at least 2400 years ago.

The builders of the polygonal walls in Italy have *not* been identified with certainty.

A long standing debate opposes those who think that most of the walls were built by the Romans or by Latin people but under the strict and direct influence of the Romans, and therefore date the most ancient of these huge constructions not before the fourth century BC, and those who think that Romans simply had nothing to do with such buildings and therefore that the walls can be several hundreds of years older than this.

The debate on the builders of the Italian polygonal walls was active in the beginning of the 20 century, when most scholars adhered to a long-standing tradition which attributed the construction of the walls to a people called *Pelasgi*, who allegedly come to Italy during the end of the Bronze Age (1200-800 BC) importing the Hittite-Mycenaean technology. However, proofs of the existence of this people are lacking and, as a matter of fact, the diffusion of techniques need not to be identified with the diffusion of a population bringing it, since a technique can be imported *via* cultural and economic contacts or can be independently invented (it suffices to think to the magnificent polygonal walls built by the Incas two thousands years later). Thus, the walls might be attributed as well to the populations inhabiting the areas before the Roman expansion, people who were, like their contemporary Etruscans who lived in adjacent areas, in active economic and cultural contacts with the civilizations of the Mediterranean sea. In 1957 however, this explanation was refused by the authoritative scholar Giuseppe Lugli[2] who re-formulated, this time as an apparently unassailable dogma, the aforementioned idea that the polygonal walls were nothing but one of the types of roman walls, a type called *by him* – I repeat, *by him* - *opus siliceum,* to be *added* to the already well known types of roman walls, such as for instance *opus quadratum*, walls of stone blocks cut as parallelepipeds and set in place on horizontal layers without mortar, and *opus caementicium*, in which the core of the walls is made of pieces of stones mixed with mortar and sand, while the external faces are of small stones or bricks.

Recently, the position of many scholars on the builders of the polygonal walls is again becoming more dubitative[3] and I deserve to deal at depth with the important problem of the dating of the walls in a separate publication. However, as we shall see, the dating of the Alatri walls is of fundamental importance for the present paper, since the results support a pre-Roman construction. Therefore, a discussion of the topic is necessary.

## 2 On the dating of Italian polygonal walls

Dating of stone building *per se* is impossible. While it is common to find on Roman bricks or even on Roman square blocks the presence of marks and timbres identifying the "producers" of the material and in many cases also the year of its extraction from the cave, no traces of this kind have ever been found on polygonal blocks anywhere in Italy. Actually Lugli's dogma of a Roman dating was founded only on a very debatable interpretation of a statement of the roman architect Vitruvius[4], who *cites* the existence of a sort of construction called *silice* without actually describing it, and on a even more debatable interpretation of a short inscription (dated around 140 BC) which runs on the walls of the Ferentino acropolis. In this inscription, the *censors* (magistrates) who certainly built the upper part of the construction in square blocks, where the inscription is actually located, credit themselves also of the construction of the lower part of the same building, in polygonal blocks, calling it, again, *silice*. The text is, however, quite ambiguous.[5]

Against Lugli's *dogma* many points can be raised, perhaps the main being the following:

1) Romans were, of course, skilled in working with huge objects, such as the granite columns of the Pantheon in Rome, weighting more than two hundreds tons and coming from south Egypt. They were skilled in building with huge blocks as well, as one can easily check visiting, for instance, the internal rooms of St. Angelo Castle in Rome, which were originally built as the funerary chambers of the Mausoleum of the emperor Adriano. However, *all* the stone buildings *in Rome* were constructed with square blocks (*opus quadratum*) from the very beginning of the Roman civilization, as one can see, for instance, from the remains of the archaic defensive walls of the town, on the *Palatino* hill, dated around 530 BC, or from the internal walls of the prison called *Carcer Tullianum,* which were built between the 6 and the 5 century BC, not to mention the first complete defensive walls of Rome, called *serviane*, dated around 390 BC. The fact that polygonal walls were never built in Rome is sometimes explained with the observation that the quality of stone most easily available *near* the town, called *Tufa*, is not easily cut in polygonal blocks.[6] If this argument can perhaps explain the most ancient constructions in Rome, it is, at least in my opinion, really flawed insofar as the civilization which was going to become the owner of the whole western world is concerned.
2) The megalithic style of the polygonal walls (with blocks that can be as heavy as 30 tons) requires techniques which are completely different from those of *opus quadratum*, in which even the hugest blocks reach at most the weight of a few tons. Romans made large use of tackles and pulleys, but stones like those visible in Alatri, in Segni and in many other places can be efficiently raised only with the help of earth ramps (perhaps one can raise *one* stone of this kind with the help of an appropriate system of pulleys, but building kilometres of walls with such devices looks impossible). In addition, before putting third-manner polygonal blocks *in situ* a hard work of fine-cutting is required, in order to fit perfectly the angles of the stones already positioned, while square blocks can be used exactly as they arrive from the cave. Once again, it is frankly difficult to believe (or at least, it is difficult to the present author) that such a spectacular and sophisticated technique was deserved to the province.
3) Techniques based on huge blocks were used by the Romans for the restraint walls of earthworks in sub-urban "Villas" (factories) of republican age (II-I century BC). However, such walls cannot be classified into one of the three manners and in fact in order to describe them Lugli was obliged to introduce an *ad hoc* a "fourth type" described as "very near to the square blocks technique". In the walls constructed with this "fourth manner" - actually a sort of "missing ring" of Lugli's dogma - there is a clear tendency to the use of horizontal setting layers, although the disposition is inaccurate and some blocks are trapezoidal and exhibit acute angles. This style is actually *the unique* used in most, if not in all, the foundations of "Villas" and, as a matter of fact, rather than considering it an "evolved" polygonal blocks technique it could be safely (and very reasonably) interpreted as a naïve way of building with square blocks, which was adopted because the construction did not require aesthetical beauty (for instance, the walls were used for agricultural terraces). The construction technique of the walls of the Acropolis in Ferentino could perhaps be included in this "fourth manner" as well, thereby accepting the credits of the two censors.

4) If the Romans used such complicated megalithic techniques, it is strange that they did not leave any document describing or at least depicting it, even in disguise. There is no Roman historian citing them, including Livius who described the foundation ("deduction") of many colonies where, according to Lugli's dogma, polygonal walls were constructed immediately thereafter the setting of the colonists. We do have many stelae illustrating buildings or even construction techniques of buildings, but in *all* such documents the blocks shown are regular, squared blocks.

5) Another point which, at least in my view, shows how much the "Roman dogma" is far from being the truth is the following. The megalithic builders did not use the arc: their "arcs" are trilithon doorways or the so called "false arcs" (a very bad terminology) composed by corbelled blocks forming a "V" upside down. The Roman architects instead used the arc from the very beginning of the Roman civilization. As a consequence, one can see Roman restoration and integration of polygonal walls in which previously existing "false arcs" were substituted by "true arcs".

As we have seen, those people who refuse Lugli's *dogma* (people to which I do belong) usually think that the construction of most of the walls pre-dates the Romans and can be attributed to the populations inhabiting the area between the VIII century BC and the Roman expansion in the IV century. Actually however only a few sites have been dated so far in a secure way (with the use of organic material or pottery associated to the walls) and some of them turned out to belong to such a period.[7] Further archaeological work in this direction is certainly advisable and the old idea of a Bronze Age retro-dating of the walls cannot, at present, be ruled out completely. As a matter of fact, it already happened in the history of archaeology that the dating of a stone building had to be shifted back in time of as much as one thousand years (I am, of course, referring to the Sarsen phase of Stonehenge, today dated around 2100 BC [8]).

## 3 The Alatri Acropolis

The city of Alatri was built around a small hill, and the town is surrounded by megalithic walls of which many remains are still visible today. The Acropolis in turn is a gigantic construction built on the hill and covering the top of it. In some sense the hill was adapted and sculpted in such a way as to obtain a sort of "geometric castle" dominating the center of the town (Fig. 1). The Acropolis is so impressive that the famous German historian Gregorovius (1821-1891) reported "an impression greater than that made by the Coliseum".
The perimeter of the building is defined by huge walls which give to it a polygonal shape with six sides ABCDEF (Fig. 2). The polygonal shape of the Acropolis actually looks like a giant replica of one of the stone blocks of which the Acropolis itself is constructed. The access to the Acropolis was possible trough two doors, today called *Porta Minore* and *Porta Maggiore*. Porta Minore is a small trilithon doorway (Fig. 3), and on its lintel a symbol composed by three phalli disposed as to form the upper part of a crux can be discerned (Fig. 4). Porta Maggiore is one among the most magnificent megalithic structure of Europe, and it is composed by a tunnel of huge stone blocks with corbelled ceiling and a monumental trilithon access (Fig. 5). After the door on the C wall, a short bent wall leads to side D which contains three huge "niches". These "niches" look as basements for statues (Fig. 6), however no kind of artistic find has been found here or elsewhere in the Acropolis. Thus, the unique "message" left today by the builders of this monument is the half-crux shaped phallic symbol on the small door (a statue of a lion, badly damaged, lies on the ramp on the north side, but the dating of this sculpture is unknown).

On the top of the Acropolis a further megalithic structure, lying on a natural rocky platform, existed. Archaeologists call this structure *ierone*, thinking that it was the basement of a temple. However, no proof of this statement has ever been found. The structure was constructed with enormous stone blocks perfectly cut and joined, without mortar, in such a way that it is impossible to insert even a paper sheet between two blocks (Fig. 7). On this building the Alatri cathedral was later constructed, so that today only the lower courses of stones remain visible, under the church.

## 4 Solar alignments and geometrical symmetries in the Alatri construction plan.

The first to propose the idea that the city of Alatri and its Acropolis were planned on the basis of geometrical and astronomical alignments was a local historian, Don Giuseppe Capone, in 1982. His work is poorly known and of very difficult access, therefore I will review here the most important discoveries of this scholar. [9]

Capone studied the geometry of the city plan and discovered that there exist a sort of "privileged point" (indicated by O in the figures 2 and 8) which lies just behind the northern wall of the megalithic structure at the center of the Acropolis (and thus it lies today near the northern wall of the Church, se again Fig. 7). With respect to the point O, Capone individuated many geometrical symmetries and astronomical alignments (Fig. 8):

1) The H corner at north-east of the acropolis defines a direction OA which individuates the rising sun at the summer solstice.
2) The eastern and western sides of the Acropolis are parallel and oriented north-south
3) The city has six main doors (indicated by P1-P6) and three small doors or *portelle* (p1-p3). The north-west sector has two main doors and two small ones, the north-east sector two main doors. "Symmetrically dividing by two" with respect to O, the south-west sector has one main door and the south-east sector one main door and one small door.
4) The lines connecting the couples of doors p1-p3, p2-P4, P3-P5 intersect each other in O.
5) The p1-p3 line is orthogonal to BA .
6) All the main doors excluding P2 are equidistant from O. The distance equals 3 times the value of the segment OH (which is about 92 meters long). This value seems to have governed the whole planning of the city

Capone was also puzzling about the shape of the Acropolis, which is *not* governed by the morphology of the hill. As a matter of fact, it is rather the contrary, because the hill was adapted, sculpted in such a way to obtain the desired shape. The idea that Capone courageously suggested was, that the plan of the Acropolis could have been conceived as an image of the Gemini constellation.

It should be noted that Capone's idea was without any doubts of great originality. In fact today we do know many examples which show how people of the past tried to connect earth and heavens, and even to replicate the sky on the ground, by means of astronomically related buildings. Some such examples are controversial and not all scholars accept them, such as, for instance, the theory which interprets the disposition of the three Giza main pyramids as a representation of the three stars of Orion's belt[10], but other are certain, like e.g. the use of *hyerophanies* , "sacred machines" which were activated by specific celestial events; among them, the famous *Castillo* of Chichen Itza', in the Yucatan, a toltec-maya pyramid which was constructed in such a way that a light and shadow serpent descends its staircase at the equinox.[11] Another example is that of the megalithic temples of Malta. These huge buildings, constructed between 3500 and 2500 BC, were planned according to a complex cosmographic concept, which included the "shape" of the so called "mother goddess" (a feminine "fat" deity) in the internal layout of the temples, the orientation of the main axis to the rising of the Southern Cross-Centaurus asterism[12] and probably also the orientation of the left "altar" of the temple to the winter solstice sunrise.[13] Recently, the present author proposed a similar "cosmographic principle" for the Inca capital, Cusco; according to this proposal, the city could have been laid out as a *replica* of a dark cloud constellation having the form of puma.[14]

## 5 Stellar alignments of the Acropolis

I will present here the results of very preliminary investigations on the possible astronomical content of the planning of the acropolis. These results have been obtained using the material kindly provided to me by the Culture Office of the Alatri Administration Municipality and do *not* take into account possible errors in the maps as well as possible inhomogeneities and impediments of the skyline at the horizon. Therefore, the proposals have to be considered as tentative, and I am currently organizing a fieldwork in order to check their validity.

The idea that the Alatri Acropolis could have been governed by stellar alignments came to me considering the fact that the eastern and western side are oriented cardinally (according to the map, they deviate of 0.5 and 0.8 degrees west of north, respectively), and that the first available "angle" for a sight line is oriented to the summer solstice sunrise.

Why, then, they did not construct the southern side with squared angles as well? And, why they did not use squared angles, or a single bent line, for the northern side? As we have seen, the Acropolis was not constructed with a form adapted to the shape of the hill, and therefore geo-morphological reasons have to be excluded.

The orientations of the sides are the following:

1) The two southern walls are oriented about 4.7 degree west of south (C side) and about 3.4 degrees east of south (D side).
2) The two northern walls are oriented about 28 degrees west of north (E side) and about 12.6 degrees west of north (F side).

To search for an astronomical answer to this puzzle I will try to consider the scant traces which the builders left to us.

There are only two exceptions to the complete, frightening silence left by them, and both have something to do with a symbolism in which the number three played some role: three phalli forming the "half crux" on the Porta Minore, and three huge niches on wall D.

Of course, there is a very important Crux in the sky, the group of stars which from the 16 century is called Southern Cross constellation. Today, due to precession, it is invisible in my country, but these stars are currently in the lower part of their precessional cycle and they were actually visible, in the Mediterranean area, in ancient times. The importance of the asterism composed by the bright stars of Centaurus and the stars of Crux is very well documented not only in the already cited megalithic temples of Malta (3400-2000 BC) but also in the megalithic sanctuaries of Minorca[15] (around 1500-1000 BC). In addition, it has recently been found also in the orientation of the Sardinian towers called Nuraghes (around 1200-800 BC).[16]

If we take a look at the southern sky in Alatri at a reference date of - say - 400 BC we discover that Crux was only partly visible. In fact, the lower star of Crux, Acrux, culminated below the southern horizon and therefore it was not visible: the Crux actually appeared as a "half cross" composed by *three* stars (actually four if one considers also the star *epsilon-crucis*). Moreover, of the two brightest stars of Centaurus, Hadar and Rigel, only Hadar was still visible at that times.

The azimuth at rising and setting of all such stars cannot of course match the directions very close to due south defined by the aforementioned azimuths of the southern sides of the Acropolis (4.7 degree west of south and 3.4 degrees east of south). However if we consider *at the same time* both such alignments and suppose that the side containing the three niches was oriented to a position of Crux, then it can be seen that when the southernmost star of Crux had an azimuth of about 4.7 degrees west of south, the bright star Hadar had an azimuth of 3.4 degrees east of south. Therefore, I suggest that the southern sides of the Acropolis could have been oriented to the Crux-Centaurus asterism in a position when the stars were in the configuration depicted in Fig. 9. Of course due to the slowness of the precessional drift and to the scarce precision of the measures used here, the date has to be taken only as extremely indicative. Since the star *Acrux* became invisible at the Alatri latitude around 700 BC, this latter date has to be considered as a *post-quem* term for the present theory, which therefore puts the planning of the Acropolis between the VII and the IV century BC. Since however this kind of interest for the southern sky is not documented by the Romans, this astronomical orientation of the southern sides of the Acropolis, if confirmed, will add the builders of Alatri to the widespread tradition of astronomical observations of the Crux-Centaurus group of the Mediterranean area in the Bronze Age, in the meanwhile proving a pre-Roman dating of the acropolis.

We now turn to the northern sky. Looking in the region around 30 degrees west of north, one immediately sees that Capella, a bright star whose importance is well documented in many cultures throughout the world, was setting at about 32 degrees west of north as viewed from Alatri in 400 BC. The bright stars Vega and Deneb also had setting azimuths in the same region (around 34 degrees for both) however Capella seems preferable, both because it is the star nearest to the desired direction and because of the following observation.

As I have mentioned, the Italian scholar Capone noticed a similarity between the plan of Alatri and the constellation Gemini, and proposed that this could have inspired the planning of the site. Actually if we look at the northern sky in Alatri at our reference date of 400 BC, we discover that while Capella was setting, *also Gemini was setting* and this constellation was standing as a "polygon", actually very similar to the shape of the Acropolis, with the side "looking" at Capella in correspondence with the north-west "bent" side of the Acropolis (Fig. 10).

**Acknowledgements**

I am indebted with dr. Antonio Agostini, director of the Library of the Alatri Administration Municipality, for his kind help in providing the material on which this paper is based, and with Prof. Gian Luca Gregori of the University of Rome for his kind help in providing references on the Ferentino inscription. A warm thank goes also to dr. Roberto Giambo' for many discussions and for his kind help in determining the data from the maps with the best accuracy.

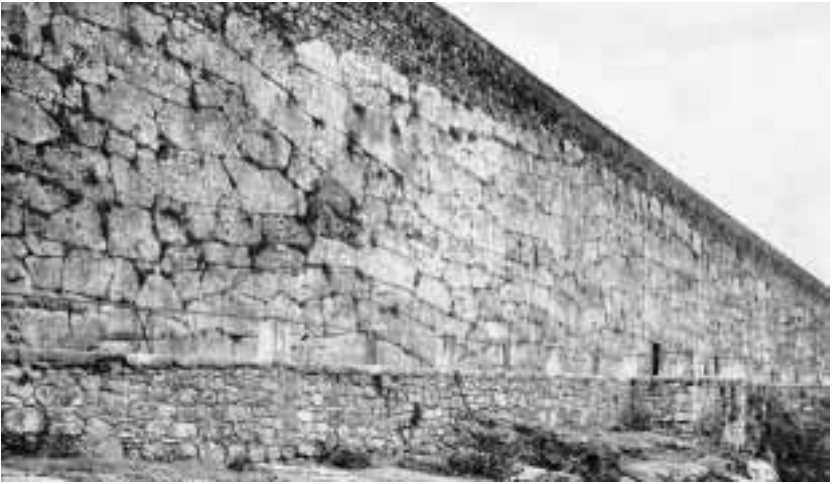

Fig. 1 Polygonal walls of the Alatri Acropolis on the side of Porta Minore, visible on the right

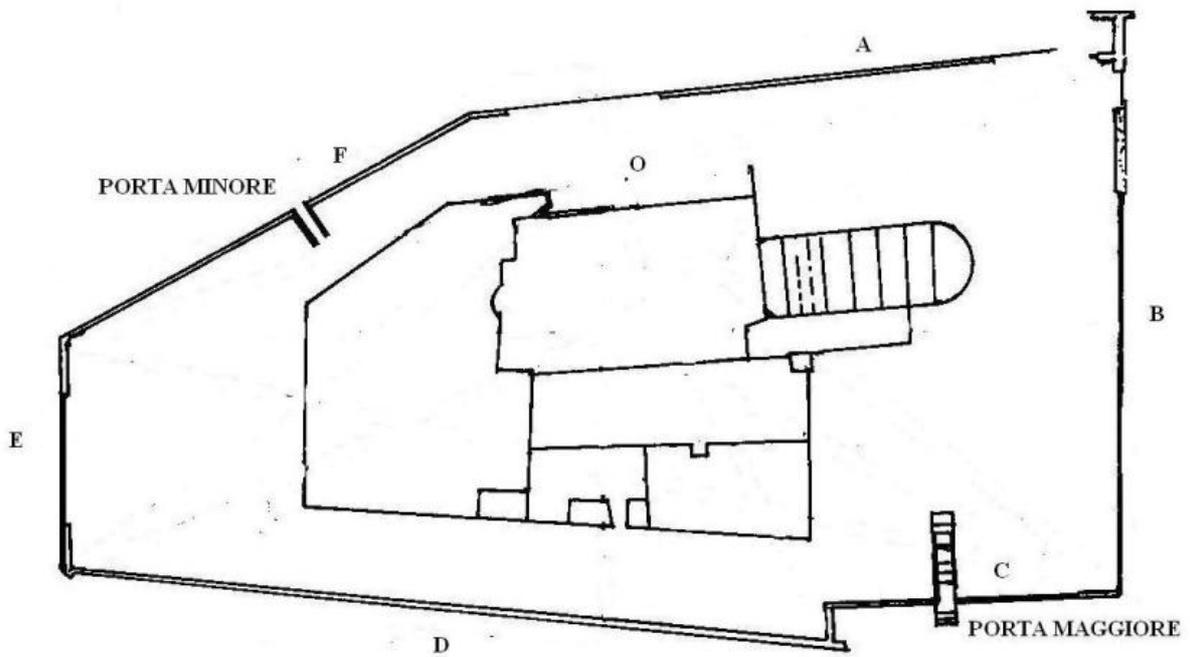

Fig. 2 Plan of the Acropolis in Alatri

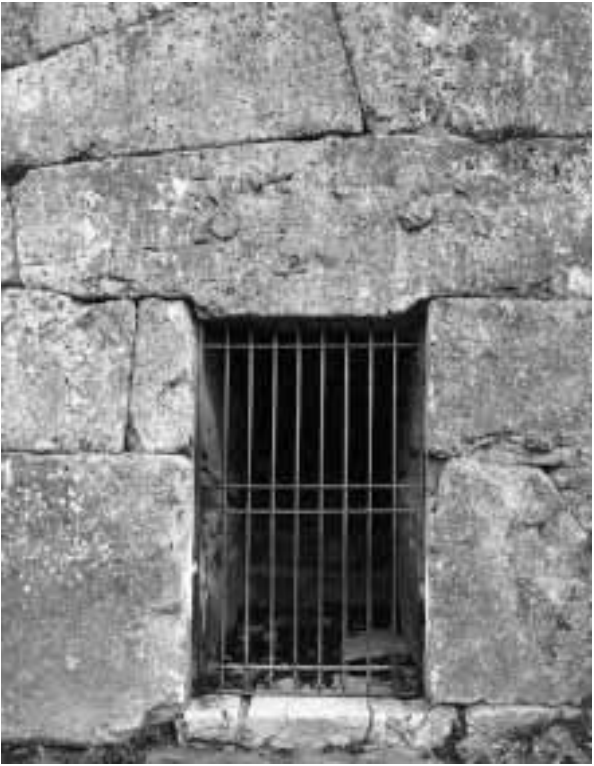

Fig. 3 Porta Minore

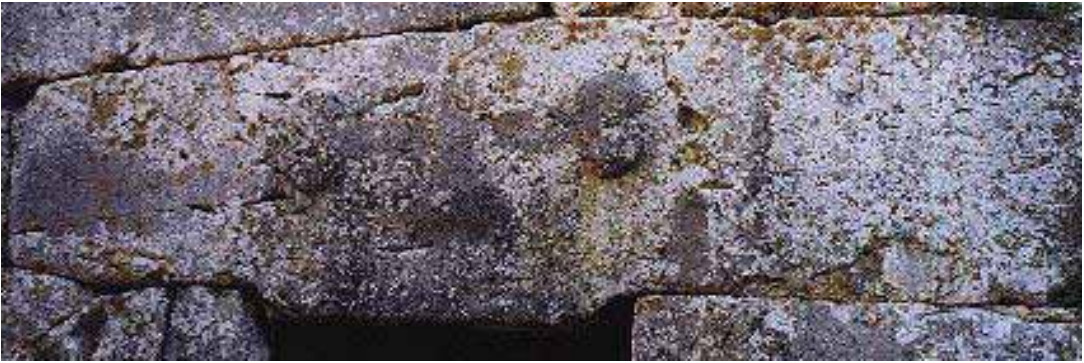

Fig. 4 Porta Minore, detail of the half crux phallic symbol

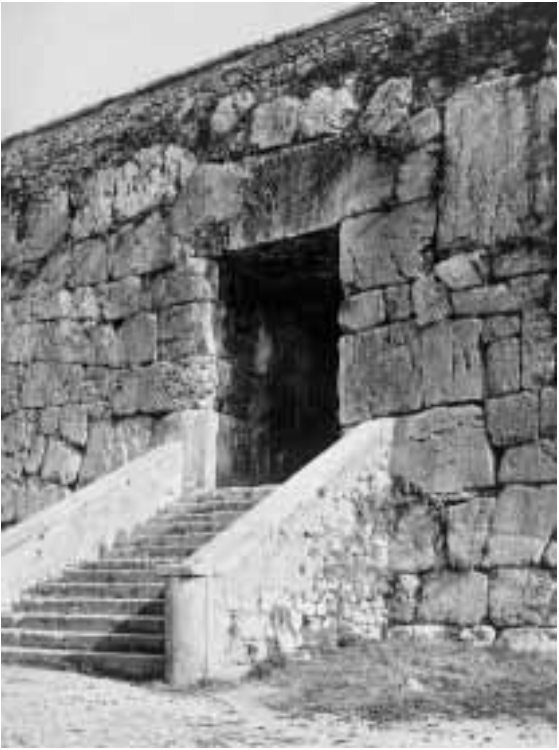

Fig. 5 Porta Maggiore. The staircase was added in the 18 century.

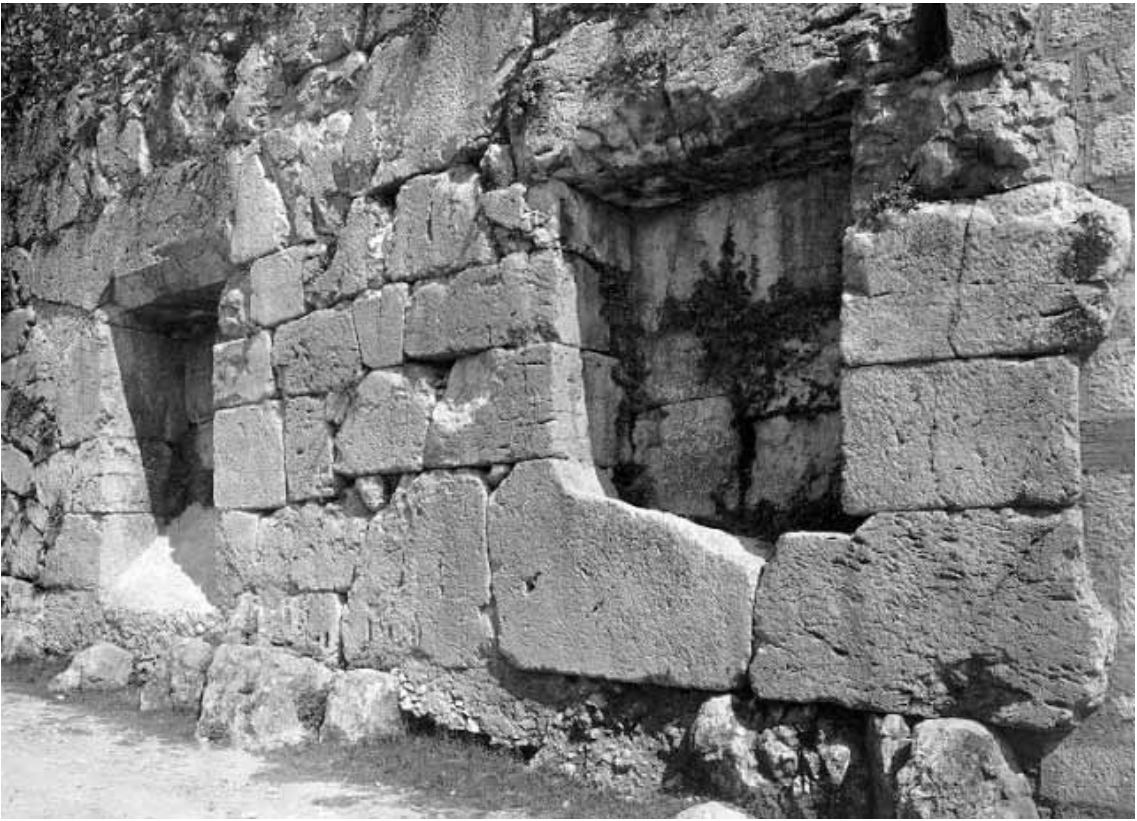

Fig. 6 Two of the three "niches" on wall D

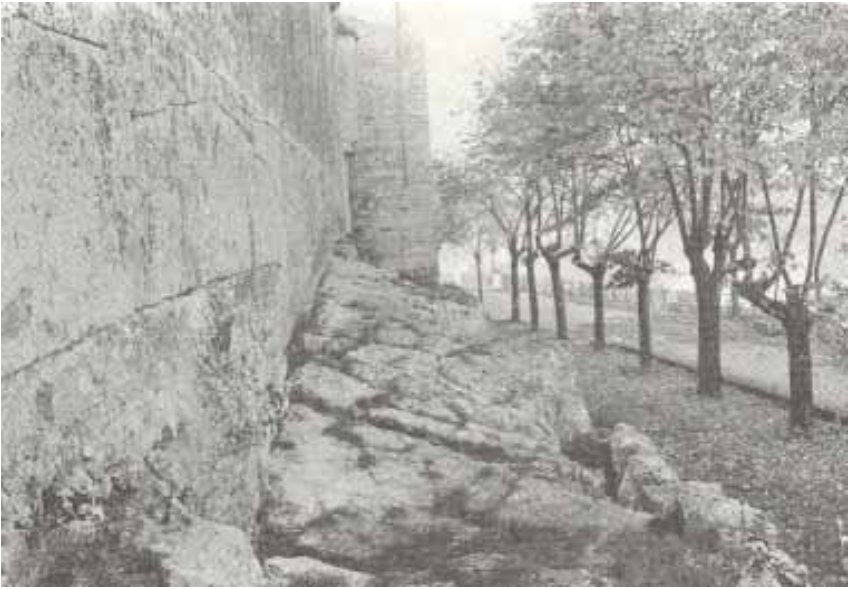

Fig. 7 Polygonal walls on the rocky terrace, visible on the northern side of the Cathedral.

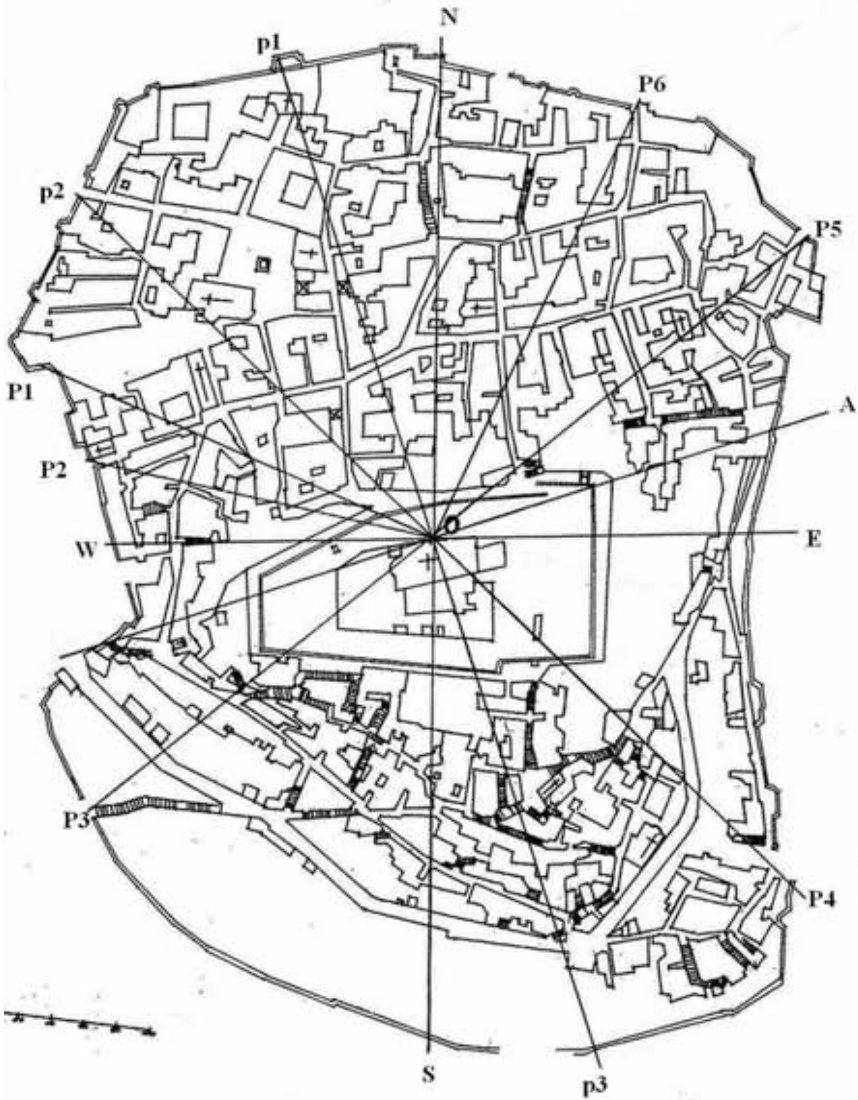

Fig. 8 Plan of Alatri, with the alignments discovered by G. Capone (adapted from Capone 1982)

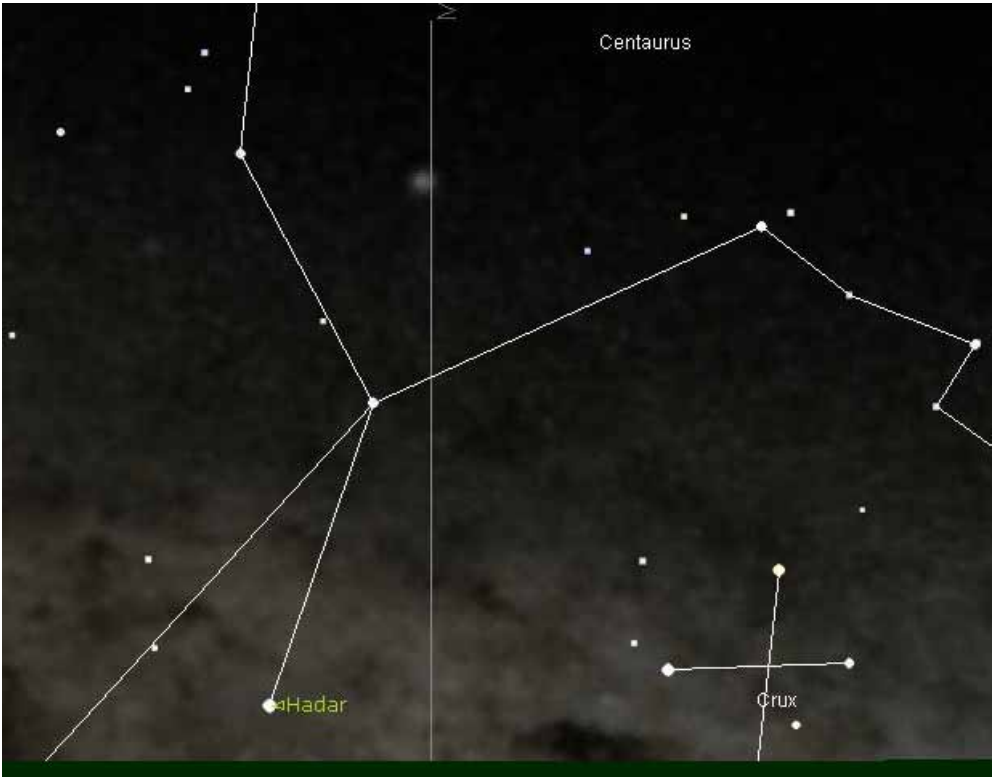

Fig. 9 The sky close to due south in Alatri in 400 BC.

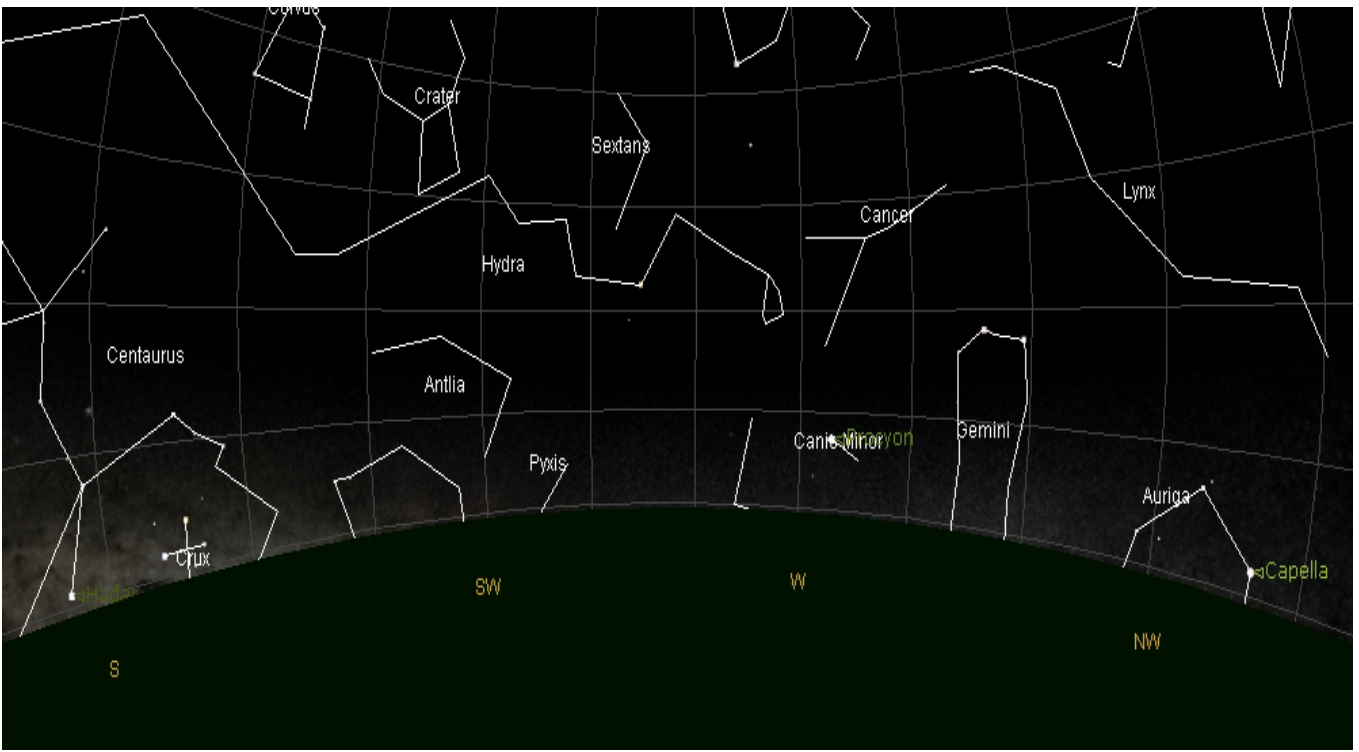

Fig. 10 Panoramic view of the sky from Alatri in 400 bc.. From left to right: on the two sides of the south pole, the Crux-Centaurus group with the bright star Hadar east of south and the Crux west of south. The lower star of Crux is always invisible (not only in this picture) due to precession. At the north-west horizon, we see Gemini in a configuration very similar to the Acropolis Plan. The "bent" side of Gemini "looks" at the bright star Capella setting at about 30 degrees west of north.

[1] See e.g. Heizer, R.F. (1990), *The age of the giants,* Marsilio-Erizzo, Venice

[2] Lugli, G. (1957) *La tecnica edilizia romana*, Bardi, Roma.

[3] See e.g. R. Marta, (1990) *Architettura Romana - Roman Architecture*, Kappa, Rome, who states "It is a characteristic of Italic civilization before Roman period: it goes from the end of the VI century to the I century BC".

[4] Vitruvius Pollio, De Architectura, 1.5.II

[5] H. Solin, (1980-82) *Le iscrizioni antiche di Ferentino*, Rend. Pont. Acc. Rom. Arch. **53-54**, p-102 and references therein.

[6] See e.g. F. Cairoli Giuliani, (1988) in *Mura Poligonali: I Seminario Nazionale di Studi*, Alatri

[7] One the few sites which have been securely dated is the series of megalithic terraces located in the site of *Norba-Monte Carbolino*, which were constructed in the VI century BC.

[8] See e.g. Pitts, M, (2001). *Hengeworld*. London: Arrow.

[9] Capone, G., (1982) *La progenie hetea*, Alatri.

[10] Bauval, R., (1989) A master plan for the three pyramids of Giza based on the three stars of the belt of Orion in Disc. Egypt. **13**, p. 7-18;

[11] Aveni, A. F. (2001). *Skywatchers: A Revised and Updated Version of Skywatchers of Ancient Mexico, University of Texas Press, Austin.*

[12] Hoskin, M. (2001) *Tombs, temples and their orientations,* Ocarina books

[13] Albrecht, K, (2001) *Maltas tempel: zwischen religion und astronomie*, Naether-Verlag, Potsdam

[14] Magli, G.,On the astronomical content of the sacred landscape of Cusco in Inka times (http://it.arxiv.org/abs/physics/0408037)

[15] See Hoskin, cit., Note 13.

[16] Zedda, M., Belmonte, J. , (2004) On the orientation of sardinian nuraghes: some clues to their interpretation. *J. Hist. Astr.* **35**, pag. 85.